\documentclass{aastex61}

\shorttitle{Sunspot number second differences as cycle precursor}
\shortauthors{Podladchikova et al.}

\begin{document}

\title{Sunspot number second differences as a precursor of the following 11-year sunspot cycle}

\correspondingauthor{Tatiana Podladchikova}
\email{t.podladchikova@skoltech.ru}

\author{Tatiana Podladchikova}
\affiliation{Skolkovo Institute of Science and Technology \\
Skolkovo Innovation Center, Building 3 \\
Moscow 143026, Russia}

\author{Ronald Van der Linden}
\affiliation{Solar-Terrestrial Center of Excellence, ROB\\
Uccle 1180, Belgium}

\author{Astrid M. Veronig}
\affiliation{Kanzelh{\"o}he Observatory \& Institute of Physics/IGAM, University of Graz \\
Universit{\"a}tsplatz 5 \\
Graz 8010, Austria}

\begin{abstract}
Forecasting the strength of the sunspot cycle is highly important for many space weather applications. 
Our previous studies have shown the importance of sunspot number variability in the declining phase 
of the current 11-year sunspot cycle to predict the strength of the next cycle when the minimum 
of the current cycle has been observed.  In this study we continue this approach and show that we can
remove the limitation of having to know the minimum epoch of the current cycle, and that we can already  
provide a forecast of the following cycle strength in the early stage of the declining phase 
of the current cycle. We introduce a method to reliably calculate sunspot number second differences 
(SNSD) in order to quantify the short-term variations of sunspot activity. We demonstrate a steady 
relationship between the SNSD dynamics in the early stage of the declining phase of a given cycle 
and the strength of the following sunspot cycle. This finding may bear physical implications 
on the underlying dynamo at work. From this relation, a relevant indicator is constructed  
that distinguishes whether the next cycle will be stronger or weaker compared to the current one. 
We demonstrate that within 24-31 months after reaching the maximum of the cycle, it can be decided 
with high probability (0.96) whether the next cycle will be weaker or stronger. 
We predict that sunspot cycle 25 will be weaker than the current cycle 24.
\end{abstract}

\keywords{Sun: activity --- Sun: sunspots --- methods: data analysis}

\section{Introduction}
Solar cycle predictions are needed to plan long-term space missions and are of high importance for space weather applications. 
Currently, precursor methods are the most favored models for the prediction of solar cycle strength 
\citep{{Conway1998}, {Svalgaard2005}, {Kane2008}, {Hathaway2009}}.
These precursor techniques often use geomagnetic activity levels near or
before the time of solar cycle minimum \citep{{OhlOhl1979},{Feynman1982},{GonzalezSchatten1988},
{Thompson1993},{WilsonHathawayReichmann1998}}.
Predicting the amplitude of a solar cycle can be done using solar polar
magnetic fields from the previous cycle as ``precursors'' of the next cycle
\citep{{SchattenSofia1987},{Schatten2005},{Svalgaard2005}, {WangSheeley2009}, {MunozJaramillo2012}}. 
The other class of precursor techniques that do not need an a priori a physical understanding of the causal
relations (i.e., that do not require any knowledge of the physics
involved) is based on finding particular sunspot number
characteristics that serve as indicators of the next cycle strength 
\citep{{Ramaswamy1977},{LantosSkewness2006},{CameronSchlussler2008},{Braja2009}}.
A number of techniques are used to predict the amplitude of a cycle during the time 
near and before sunspot minimum \citep{Hathaway2010}. 
The two precursor types that have received the most attention are polar field precursors 
and geomagnetic precursors. The strength of the solar polar magnetic fields at 
solar minimum is a very accurate indicator of the maximum amplitude of the following 
solar cycle. Forecasts using the polar field method have proven to be consistently in the right range 
for cycles 21, 22, and 23 \citep{{SchattenSofia1987},{Schatten1996}}.
The polar fields reach their maximal amplitude near the minima of the sunspot cycle. 
However, the maxima of the polar field curves are often rather flat so approximate 
forecasts are feasible several years before the actual minimum \citep{{Hathaway2010}, {Petrovay2010}}. 
Using the rather flat and low maximum in polar field strength, \citet{Svalgaard2005}
have been able to predict a relatively weak cycle 24. Such an early prediction 
is not always possible: early polar field predictions of cycles 22 and 23 had to be 
corrected later and only forecasts made shortly before the actual minimum did finally converged 
\citep{{Hathaway2010},{Petrovay2010}}. 

A physical explanation for how precursor methods work was
suggested by \citet{SchattenDynamo1978}, who used the reversed
polar field built up after the solar maximum as a precursor
indicator to the next solar cycle strength.
In the framework of a dynamo model of the Babcock -- Leighton type,
\citet{CameronSchlussler2007} have discussed the importance of
the sunspot activity level during the declining phase and have considered 
the sunspot activity three years before the minimum
as a predictor of the next cycle. 

As pointed out by \citet{CameronSchlussler2007} and discussed
by \citet{Petrovay2010}, the dependence of many precursor
methods on knowing the minimum epoch somewhat limits their practical usefulness as the epoch of the minimum of the sunspot number cannot be
known with certainty until about a year after the event. In addition, these
methods do not necessarily involve a physical connection between the
precursor and the dynamo process. The Waldmeier effect and the overlap
of solar cycles may explain the correlation of the precursor with
the amplitude of the next cycle. As stronger cycles tend to rise
faster to their maximum activity, the temporal overlap of cycles
leads to a shift of the cycle minimum to an earlier time,
when the activity from the previous cycle has not yet reached very
low levels.

In our previous studies \citep{PodladchikovaLefebvreLinden2008}, we demonstrated the importance 
of sunspot number variability during the declining phase of the sunspot cycle
to predict the peak of the ensuing cycle. The analysis revealed that a slow decline 
and short-term rise of the sunspot number during the declining phase,  
can be viewed as a presage to a stronger peak of the following cycle. 
Once the minimum of the cycle had passed, it allowed us to predict the weak 
sunspot cycle 24 with the sunspot maximum not exceeding 72 \citep{Podladchikova2011}.
The actual solar maximum of cycle 24 had a peak of 82. In that analysis we used the sunspot number, as it was before the major change of this data set on 2015 July 1 \citep{Clette2016}.

We present here a further investigation of sunspot number variations around 
the epoch of the maximum number of sunspots, a period of the strongest toroidal field,
when there is a change of magnetic polarity, and demonstrate that the evidence of a stronger or weaker
following cycle is already visible within 13-18 months after the current cycle has reached its peak.
We estimate the second derivative of the sunspot numbers
determined as sunspot number second differences (SNSD),
which are closely related with the curvature of the sunspot
series, to study the sunspot number variations.

We demonstrate a steady relationship between the SNSD variations of the current cycle 
right after reaching its peak and the strength of a following sunspot cycle, and 
construct a relevant indicator that can determine whether the next cycle will be stronger 
or weaker compared to the current one. For our analysis we use the SILSO sunspot number 
series as it is currently available following the major change on 2015 July 1 \citep{Clette2016}. 

\section{Data preparation} \label{S-var}
The sunspot number behavior is characterized not only by a complicated multi-frequency structure representing the intrinsic stochastic component of the solar cycle, but also by a (much smaller) stochastic component associated with measurement errors and systematics related to the way how the sunspot data are acquired, combining data from different instruments/observers/techniques.
To isolate the component associated with the long-term behavior of the solar cycle, 
the 13-month smoothed sunspot number is used \citep{Hathaway1999}. 
The traditional 13-month running mean, centered on a given month with equal weights for
months –5 to +5 and a half weight for months –6 and +6, is both simple and widely used but
does a poor job of filtering out high-frequency variations \citep{Hathaway2010}.
\citet{Hurst1970} examined the influence of the width of the smoothing window on the accuracy of the running mean 
and showed that the running mean may significantly distort short-term variations, while 
the variations with a period larger than the window width remain almost unchanged.  
Note that the running mean is a suboptimal estimation method that does not use 
any optimization criterion. The width of the smoothing window is in general chosen empirically,
depending on the particular problem.

In this study, we want to analyze and quantify the short-term variations of the sunspot number
time series in terms of second derivatives. Thus, we need a more reliable means to thoroughly
smooth the series while keeping the intrinsic short-term variations of the series. Thus, we propose 
a more suitable smoothing technique of sunspot numbers that is based on finding a balance between 
the fidelity to the data and the smoothness of an approximating curve \citep{Bohlmann1899, Whittaker1923, Weinert2007}.

Let $R_{i}$ denote the monthly mean sunspot number, and $\hat{R}_{i}$ be the smoothed sunspot number
at month $i$. On the basis of root mean squares we should consider the best approximation of the smoothed curve $\bar{R}$
to the monthly mean sunspot numbers $R$ if it minimizes the sum of squared deviations
\begin{equation} \label{Eq_Sd}
S_{d} = \sum\limits_{i=1}^{n}(R_{i}-\bar{R}_{i})^{2},\qquad (i=1,2,\ldots,n) 
\end{equation}
and the curve is considered to be smoother if it minimizes the sum of squared second differences
\begin{equation} \label{Eq_Sv}
S_{\nu} =\sum\limits_{i=1}^{n-2}(\bar{R}_{i+2} -2\bar{R}_{i+1}+\bar{R}_{i})^{2},\qquad (i=1,2,\ldots,n-2)
\end{equation}
The level of fidelity to the data is represented by the closeness of the smoothed sunspot curve $\bar{R}$ 
to the original monthly mean sunspot numbers $R$. It is determined by the sum of squared deviations $S_{d}$ (deviation indicator, Equation~(\ref{Eq_Sd})). 
The level of smoothness of $\bar{R}$ is determined 
by the sum of squared second differences $S_{\nu}$ (variability indicator, Equation~(\ref{Eq_Sv})).

To find the best approximation of the sunspot cycle for our purposes, we optimize the smoothing algorithm of monthly sunspot numbers
by determining the smoothed values $\bar{R_{i}}$, that minimize the following functional $J$:
\begin{equation} \label{Eq_J}
J =\beta\sum\limits_{i=1}^{n}(R_{i}-\bar{R}_{i})^{2}+\sum\limits_{i=1}^{n-2}(\bar{R}_{i+2} -2\bar{R}_{i+1}+\bar{R}_{i})^{2}
\end{equation}
Here, $\beta$ is a smoothing constant that determines how close the approximating curve fits the monthly mean sunspot numbers.
If $\beta$ is increased, the smoothed curve becomes closer to the original data series,
as the level of smoothing is reduced. When $\beta$ is decreased, 
the scatter of the approximating curve decreases, but the deviation from the original data series increases.
The value of the smoothing constant usually is chosen empirically, depending on the particular problem \citep{Brown1963, Hurst1970}.
In the following text we refer to this method as the optimized smoothing technique, which means that 
it is optimized with respect to minimizing the functional $J$ in Equation~(\ref{Eq_J}),
which serves as a balance between the fidelity of the data and smoothness.

By minimizing the functional $J$ given by Equation~(\ref{Eq_J}) and finding its derivative
with respect to $\bar{R_{i}}$, $(i=1,2,\ldots,n)$, we obtain a system of $n$ normal equations 
with $n$ unknowns, which can be presented in the following matrix form:
\begin{equation} \label{Eq_SofEq}
  A \bar{R}=\beta R
\end{equation}
Here, $A$ is the $n\times n$-dimensional five diagonal banded matrix 
\begin{equation} \label{Eq_A}
	 A = \left|
	\begin{array}{ccccccc} 
	 1+\beta & -2       &  1       &  0      &  0       &  \cdots  &  0       \\
	-2       &  5+\beta & -4       &  1      &  0       &  \cdots  &  0       \\
	 1       & -4       &  6+\beta & -4      &  1       &  \cdots  &  0       \\
	 \cdots  &  \cdots  &  \cdots  &  \cdots &  \cdots  &  \cdots  &  \cdots  \\
	 0       &  \cdots  &  1       & -4      &  6+\beta & -4       &  1       \\
	 0       &  \cdots  &  0       &  1      & -4       &  5+\beta & -2       \\
	 0       &  \cdots  &  0       &  0      &  1       & -2       &  1+\beta \\
	\end{array}\right|
\end{equation}
$R = \left|R_{1},R_{2},\cdots,R_{n}\right|^{T}$ is the $n$-dimensional vector of the monthly mean sunspot numbers, 
and $\bar{R} = \left|\bar{R}_{1},\bar{R}_{2},\cdots,\bar{R}_{n}\right|^{T}$ is the $n$-dimensional vector of smoothed sunspot numbers.
The solution of Equation~(\ref{Eq_SofEq}) is found by inverting the matrix:
\begin{equation} \label{Eq_Sol}
  \bar{R}=\beta A^{-1}R
\end{equation}
The smoothed values of $\bar{R}_{i}, (i=1,2,\ldots,n)$ are thus determined at every point $i$ as 
a weighted sum of the monthly mean sunspot numbers $R_{i}$. The weights follow from the application
of Equation~(\ref{Eq_Sol}) and are determined by the elements of row $i$ of matrix $A^{-1}$ 
multiplied by the smoothing constant $\beta$. The monthly mean sunspot number $R_{i}$ is taken into account with the maximal weight,
and the weights monotonously decrease with distance from month $i$. 
	
\begin{figure}  
\plotone{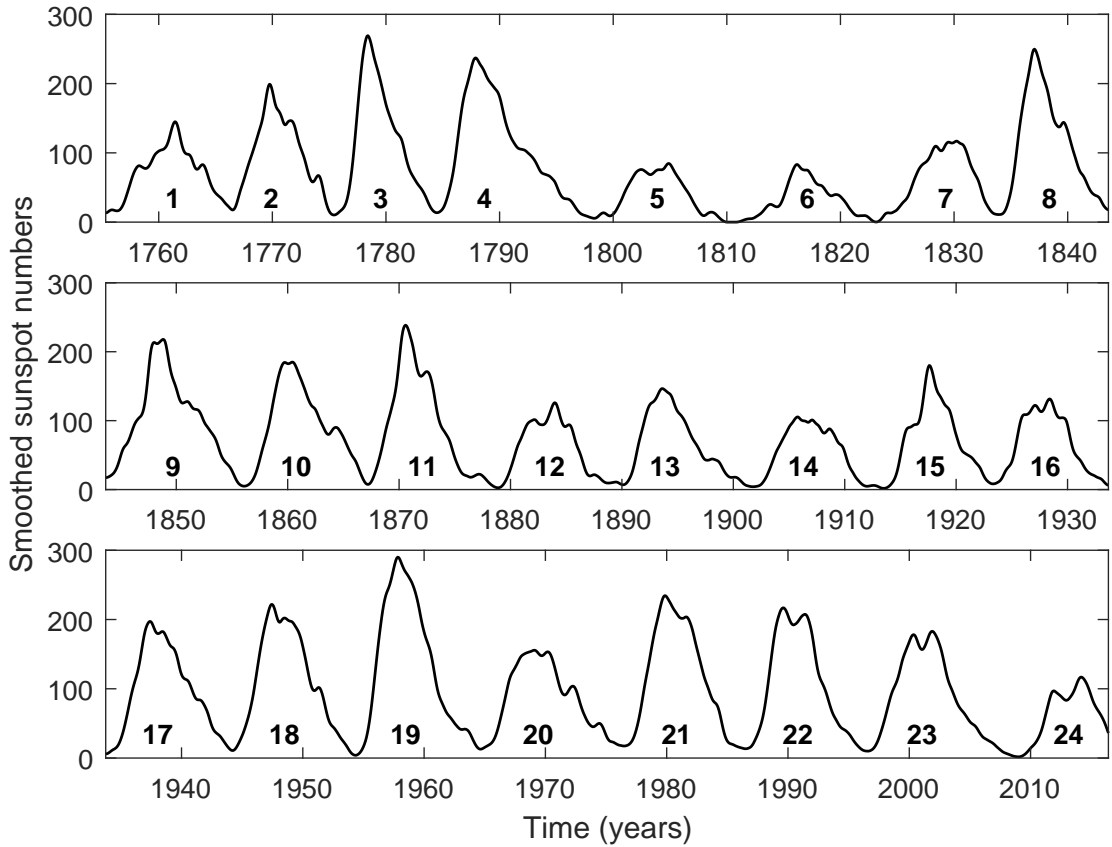}
\caption{Optimized smoothed sunspot numbers for cycles 1-24.}
\label{fig1}
\end{figure}

For our analysis we use a smoothing constant of $\beta=0.01$ that provides maximal filtration of noise,
which is a necessary preparation to reliably derive SNSD. When $\beta<0.01$, the matrix $A$ may become ill-conditioned,
which may lead to errors in the calculated smoothed sunspot series \citep{Rice81}. In the case of $\beta = 0.01$, 
the elements of the matrix that determine the coefficient weights, $\beta A^{-1}$,  become very small ($<0.001$) 
at distances $>10$ months. Thus, they effectively do not contribute to the smoothing result, and we can interpret 
the value of $\beta = 0.01$ to correspond to an effective smoothing interval of 21 months with changing weights 
for months -10 to +10. An increase of the smoothing constant $\beta$ reduces the level of smoothing, 
makes the smoothed curve closer to the original data, and shortens the smoothing interval. 
Figure~\ref{fig1} shows the smoothed sunspot numbers for $\beta=0.01$ 
for all cycles 1-24 using all the available monthly mean sunspot numbers until 2016 June. 
As shown in Figure~\ref{fig1}, many sunspot cycles exhibit double peaks, which can reflect 
significant variations in solar activity on time scales of one to three years \citep{Hathaway2010}. 
The double peaks of recent cycles 22, 23, and 24 demonstrate clear Gnevyshev gaps \citep{Gnevyshev1967}.

\begin{figure}  
\plotone{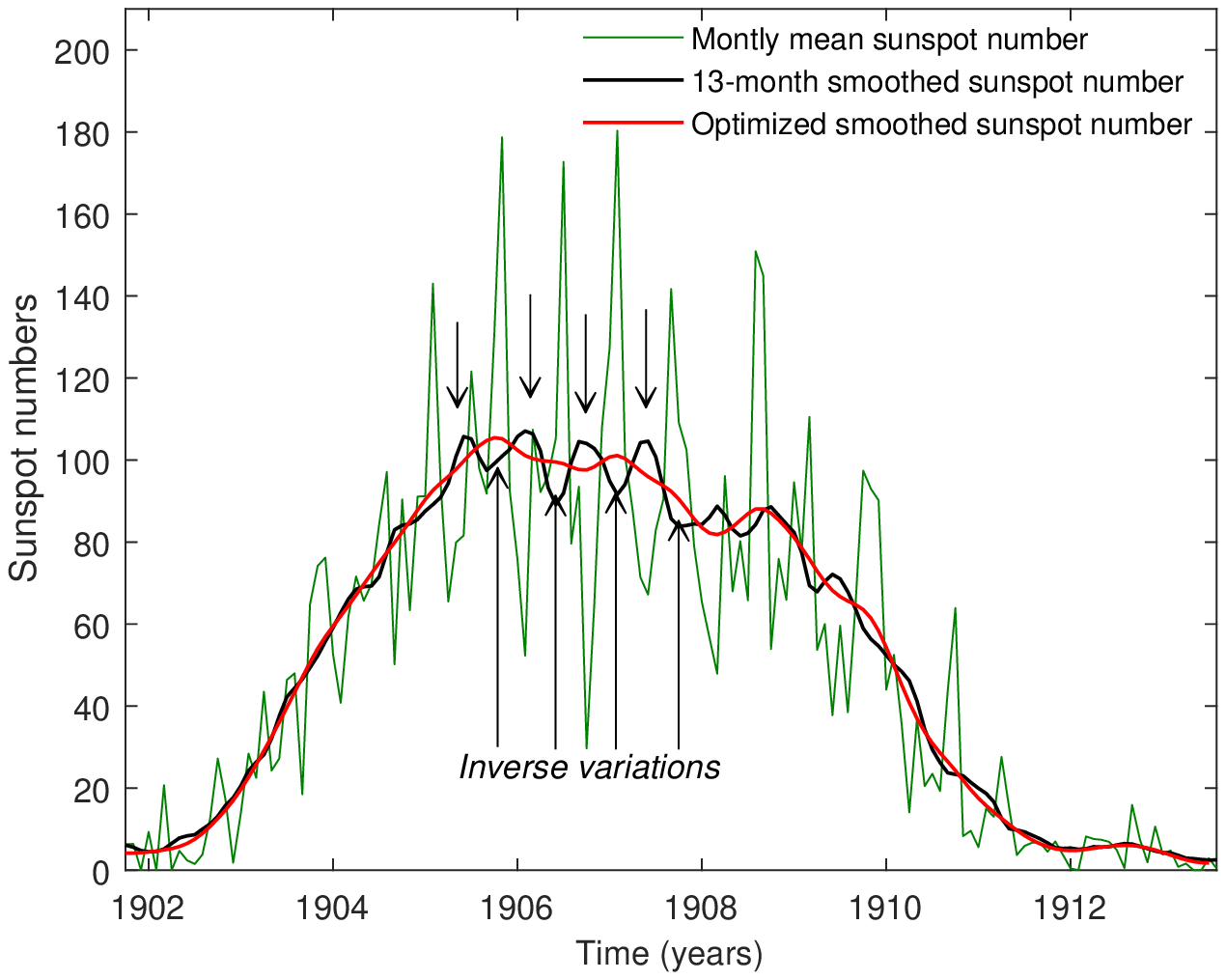}
\caption{Sunspot cycle 14: monthly mean sunspot number (green), 
13-month smoothed sunspot numbers (black), and optimized smoothed sunspot numbers (red).
The arrows indicate the areas where the 13-month smoothed (running mean) sunspot numbers produce 
inverse variations to the original time series. This effect does not appear in the optimized smoothed data.}
\label{fig2}
\end{figure}

To compare the optimized smoothing technique with the 13-month running mean, in Figure~\ref{fig2} we present both 
curves for cycle 14. The monthly mean sunspot numbers are given by the green line,
the black line depicts the 13-month smoothed sunspot numbers, and the optimized smoothed sunspot numbers
are indicated by the red line. As shown in Figure~\ref{fig2}, both filters follow the long-term dynamics of the sunspot cycle. 
However, in many cases the 13-month smoothed sunspot numbers lose traces 
of the short-term variations of the sunspot cycle. The 13-month smoothed sunspot numbers increase (black line) 
over short-term periods, in which the monthly mean sunspot numbers actually decrease (green line) and vice versa
(indicated by the arrows). Such unwanted behavior does not show up in the smoothing filter presented here.
This is a strength of the method and results from the combined minimization criteria of fidelity to the data
and smoothness in Equation~(\ref{Eq_J}).

\begin{figure}  
\plotone{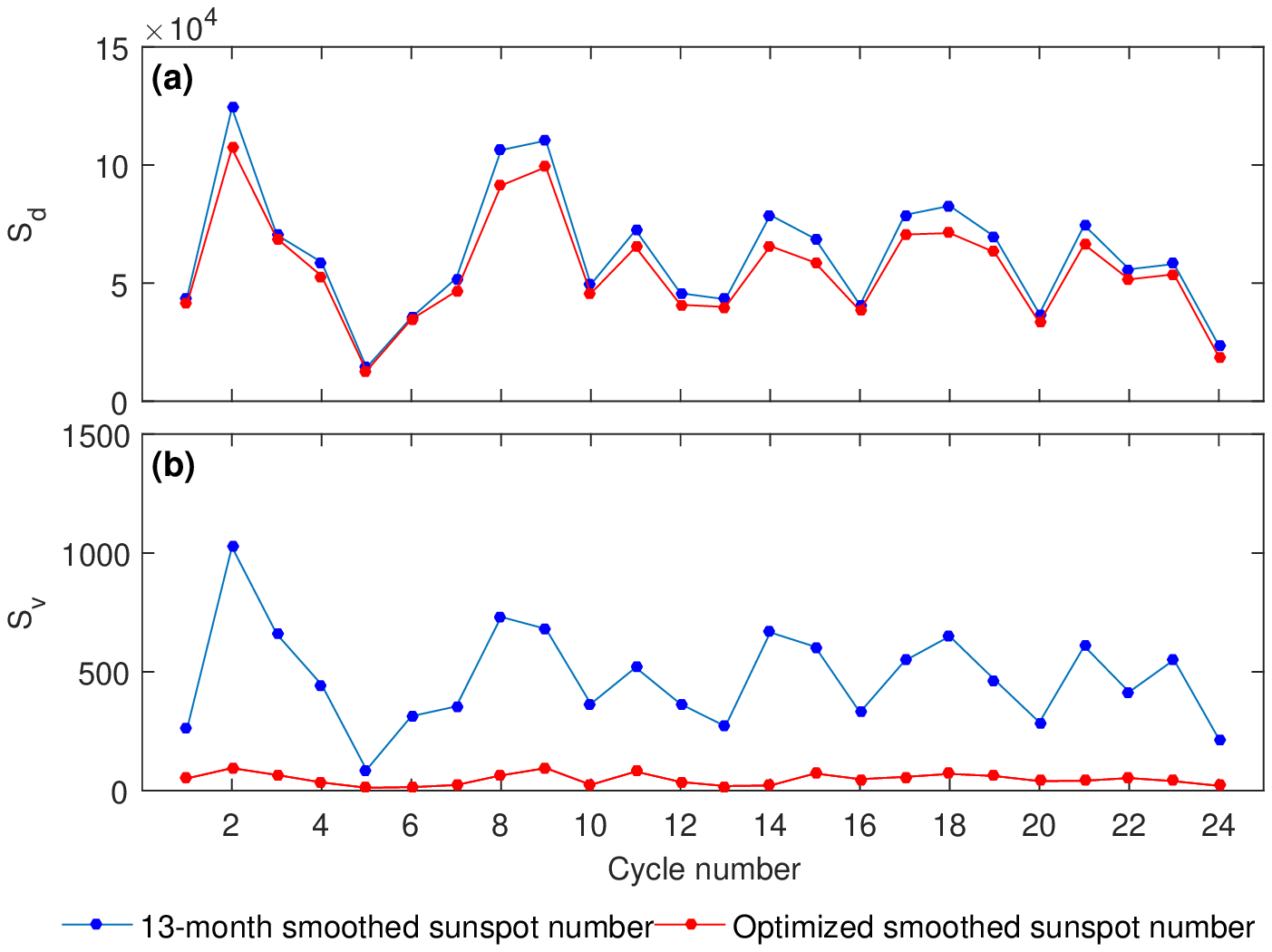}
\caption{Deviation indicator $S_{d}$ (a) and the variability indicator $S_{v}$ (b) for all sunspot cycles 1--24
derived for the 13-month smoothed sunspot numbers (blue), and for the optimized smoothed sunspot numbers (red).}
\label{fig3}
\end{figure}

To quantitatively compare the considered smoothing filters, we determine the deviation indicator 
$S_{d}$ that determines the closeness of the smoothed curve to the monthly mean sunspot numbers 
(Equation~(\ref{Eq_Sd})) and the variability indicator $S_{\nu}$ that represents the level of smoothness 
(Equation~(\ref{Eq_Sv})) for all cycles 1-24 and the results are plotted in Figure~\ref{fig3}.  
The deviation indicator $S_{d}$ is shown in Figure~\ref{fig3}a and the variability indicator $S_{v}$ is shown
in Figure~\ref{fig3}b. The blue and red lines indicate the values of $S_{d}$ and $S_{v}$ 
for the 13-month running mean and for the optimized smoothing technique, respectively.  
As shown in Figure~\ref{fig3}, the deviation indicator $S_{d}$ shows a better match of the optimized  
smoothed sunspot numbers to the original data series compared to the case for the 13-month smoothed (running mean) sunspot numbers (Figure~\ref{fig3}a). The variability indicator $S_{v}$ demonstrates the significantly smaller variability 
of the optimized smoothed curves compared to the 13-month smoothed sunspot numbers (Figure~\ref{fig3}b). 
Thus, the optimized smoothing technique provides a better compromise between the fidelity
to the data and noise filtration compared to the traditional 13-month running mean. 

\section{Sunspot number variability in the solar cycle descending phase} \label{S-var}
\begin{figure}  
\plotone{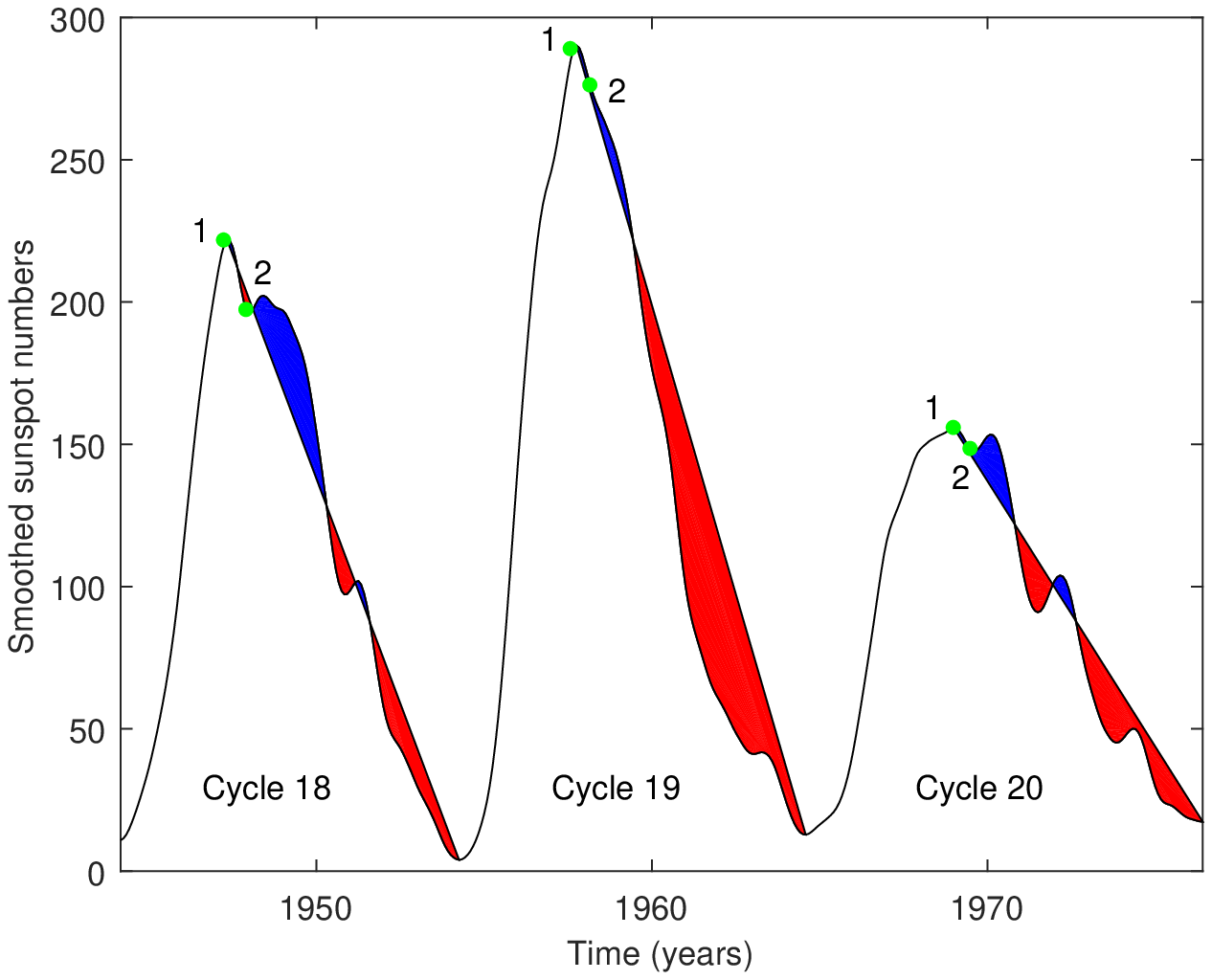}
\caption{Smoothed sunspot numbers for cycles 18, 19, and 20.
The blue-shaded areas characterize an ``excess'' activity compared to a “uniform decrease”. 
The red-shaded areas represent an activity ``shortfall'' in comparison with a uniform decline.
The green points correspond to the points $a_{min}$ (1) and $a_{max}$ (2) in Figure~\ref{fig5}.}
\label{fig4}
\end{figure}

In our previous studies \citep{{PodladchikovaLefebvreLinden2008},{Podladchikova2011}} we demonstrated 
the relation of the sunspot number variability in the descending phase of a given 11-year sunspot cycle
to the strength of the ensuing cycle. The analysis revealed that a slow decline of the sunspot number 
or rapid variations during the declining phase can be viewed as an indicator of a larger peak of the next cycle compared to the current one. For illustration, Figure~\ref{fig4} shows 
the evolution of the optimized smoothed sunspot numbers during the three consecutive 11-year cycles: 18, 19, and 20. 
The blue-shaded areas characterize ``excess'' activity compared to a uniform decrease. 
The red-shaded areas represent an activity ``shortfall'' in comparison with a uniform decline. 
The analysis of the blue-shaded areas in relation to the red-shaded ones allowed us to predict 
the strength of the next sunspot cycle once the minimum of the current cycle had been observed. 
As shown in Figure~\ref{fig4}, cycles 18 and 20, which are followed by stronger cycles, 
are characterized by larger blue areas in the early stage of the declining phase
compared to that in cycle 19, which is followed by a weaker cycle. 
On the other hand, cycle 19 has significantly larger red areas compared to those in cycles 18 and 20. 
Sudden variations of activity in the declining phase
are associated with a slowdown of the decline of the sunspot
number, which can be evidence of activity that manifests itself
as a larger amplitude in the next cycle.
Cycles 18 and 20 (followed by stronger cycles) already exhibit significant difference 
in ``excess'' activity and decline rate already in the early stage of the descending phase, 
compared to that in cycle 19, preceding a weaker cycle. In the present study, we investigate whether these
short-term variations of sunspot activity can be used to forecast the strength of the next cycle when already 
in the early stage of the declining phase of the current cycle. 

\section{Sunspot number second differences} \label{S-SNSD}
For the analysis of the solar activity variations, we determine the SNSD, a quantity that is closely related to the curvature of the sunspot series.
Rapid and abrupt variations of sunspot activity are superimposed upon the  
long-term trend of the solar cycle (see, e.g. Figure~\ref{fig4}). To remove the long-term variations in the
sunspot number record and focus only on the short-term variations, we calculate the second differences 
$\bar{R}_{i+1} -2\bar{R}_{i}+\bar{R}_{i-1}$ of the optimized smoothed sunspot numbers $\bar{R}_{i}$ from 1749 to 2016 June. 
The values of SNSD quantify how the rate of growth or decline of the sunspot numbers changes.
Negative (positive) values of SNSD indicate either a slower (faster) rise or a faster (slower) decline of sunspot activity.

\begin{figure}  
\plotone{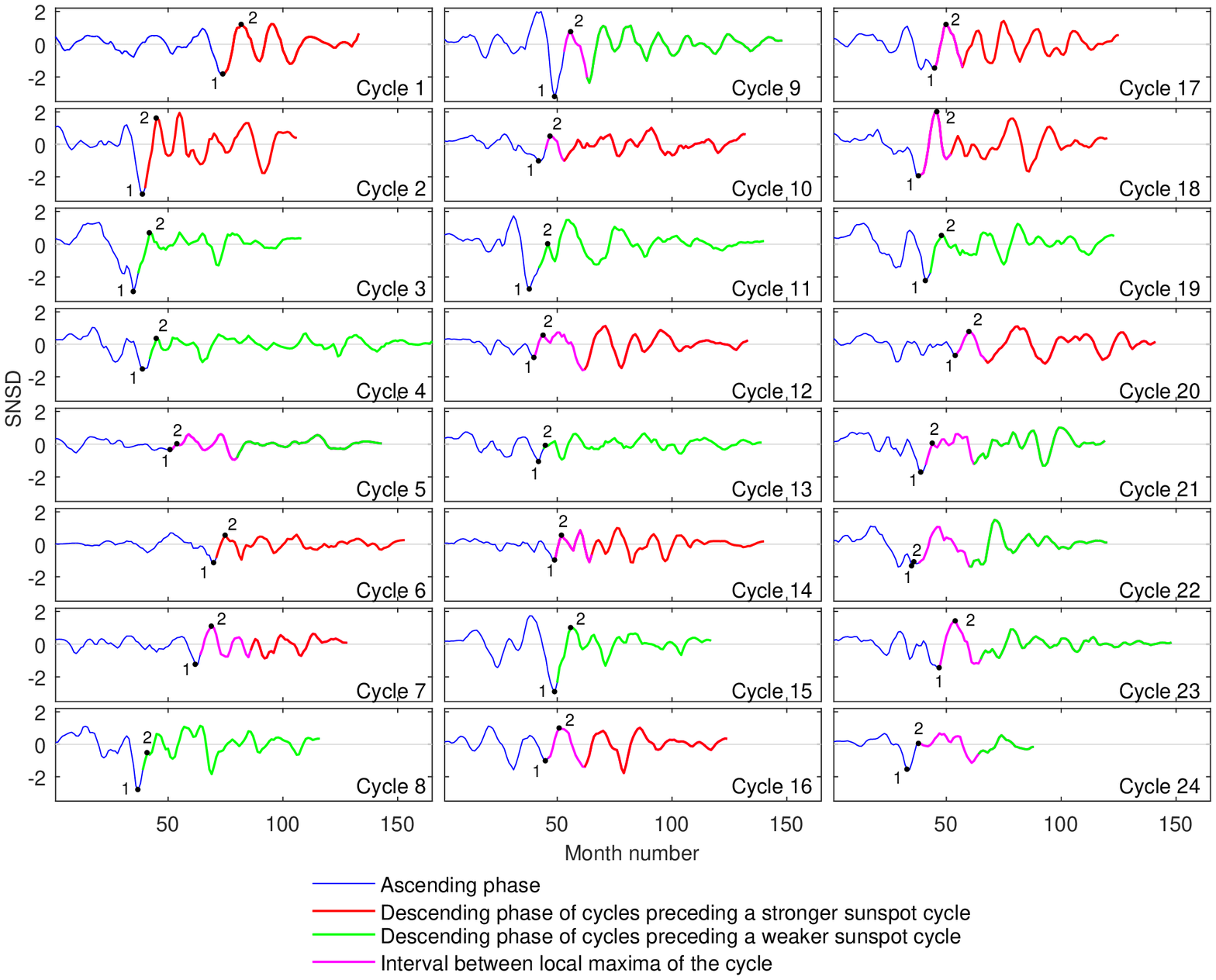}
\caption{SNSDs for cycles 1-24.
The ascending phase of each cycle to its first peak is indicated by the blue lines.
The descending phases of cycles that precede a stronger cycle are given by red lines,
whereas the descending phases of cycles that precede a weaker cycle are given by green lines.
If a cycle is characterized by two or more peaks, then the interval between
the first and the latest peak is marked by a magenta line.
Point 1 indicates the negative SNSD peak $a_{min}$ around a cycle peak.
Point 2 marks the SNSD peak $a_{max}$ corresponding to the peak of the first ensuing 
opposite curvature of the SNSD.}
\label{fig5}
\end{figure}

Figure~\ref{fig5} shows the SNSD for all cycles 1-24. 
One can see that the dynamics of the SNSD in the ascending and declining phases have a different nature
and the SNSDs in the declining phase are characterized by quasi-periodic oscillations around 
the zero level that fade out when approaching the minimum epoch. 
Around the cycle peak, the SNSDs fall to a negative peak $a_{min}$ marked in Figure~\ref{fig5} 
by the black point 1. If a cycle is characterized by two or more peaks, then the black point 1
in Figure~\ref{fig5} is defined to indicate the SNSD peak $a_{min}$ around the first cycle peak.

The registration time of $a_{min}$ on the sunspot curve for cycles 18, 19, and 20
is also indicated by the green point 1 in Figure~\ref{fig4}. After reaching the negative peak $a_{min}$ 
the SNSDs increase to a peak $a_{max}$ corresponding to the peak of the 	
first ensuing opposite curvature of the SNSD, which is marked by the black point 2 in Figure~\ref{fig5}. 
This is followed by a further slowdown of sunspot number decline or even to a short-term rise 
of the sunspot activity. 
This is shown in Figure~\ref{fig4} where the corresponding peaks $a_{max}$ are indicated by the green point 2.
Thus, the indicated SNSD oscillations and short-term sunspot number variations occur 
in anti-phase and the SNSD oscillations \textit{precede} those of sunspot numbers. 
This means that the SNSD dynamics from $a_{min}$ to $a_{max}$ is an indicator of the dynamics of the further declining phase. 

As shown in Figure~\ref{fig5}, the SNSD around the maximum of cycle 19 is characterized by 
a smaller negative peak $a_{min}=-2.25$ (point 1) and a smaller positive peak $a_{max}=0.51$ (point 2)
compared to those for cycles 18 and 20, which have negative peaks $a_{min}=-1.96$ 
and $a_{min}=-0.71$ (point 1), and positive peaks $a_{max}=1.99$ and $a_{max}=0.76$ (point 2). 
The increase of SNSD for cycle 19, from the deepest negative peak $a_{min}$ (point 1), to the smallest positive peak 
$a_{max}$ (point 2), demonstrates the weak growth process of the second derivative,
while the greater peaks $a_{min}$ and $a_{max}$ 
for cycles 18 and 20 indicate ``excess'' activity that can be considered as evidence of the existence 
of a precursor activity that shall manifest itself by a larger amplitude of the next cycle.
Similarly, cycles 3 and 4 preceding weaker cycles demonstrate relatively weak SNSD growth from $a_{min}$ to $a_{max}$
compared to cycles 1 and 2 preceding stronger cycles, and the same is true for cycle 7 with respect to cycle 8. 

Thus, the SNSD dynamics of the current cycle right after reaching its peak demonstrates its predictive power 
for the strength of the next cycle. The values of SNSD peaks $a_{min}$ and $a_{max}$ for cycles 1-24 marked 
by the black points 1 and 2 in Figure~\ref{fig5} are listed in Table~\ref{table1}. 
\begin{table}
\centering
\caption{The SNSD peaks $a_{min}$ and $a_{max}$ for cycles 1-24. The row ``Cycle'' gives  the cycle number.
The row ``Period'' gives the time in months between the peaks.} 
\label{table1}
\begin{tabular}{lcccccccc}     
  \hline                            
  \hline                   
Cycle      & 1      & 2     & 3     & 4      & 5     & 6      & 7      & 8     \\
$a_{min}$  & -1.84  &-3.09  &-2.91  & -1.54  & -0.35 & -1.15  & -1.25  &-2.81  \\
$a_{max}$  & 1.19   & 1.59  & 0.67	& 0.34	 & 0.006 & 0.53   & 1.08   &-0.54  \\
Period   	 & 8     	& 6     & 7     & 6      & 3     & 5      & 7      & 4     \\
  \hline
Cycle      & 9      & 10    & 11    & 12     & 13    & 14     & 15     & 16    \\
$a_{min}$  & -3.22  & -1.04 & -2.76 & -0.83  & -1.08 & -1.00  & -2.92  & -1.04 \\
$a_{max}$  & 0.74	  & 0.48  & 0.006 & 0.54   & -0.09 & 0.54	  & 0.99   & 0.98  \\
Period     & 7      & 5     & 8     & 4      & 7     & 3      & 7      & 6     \\
	\hline
Cycle      & 17     & 18    & 19    & 20     & 21    & 22     & 23     & 24    \\
$a_{min}$  & -1.48  & -1.96 & -2.25 & -0.71  & -1.73 & -1.1   & -1.46	 & -1.56 \\
$a_{max}$  & 1.19   & 1.99	& 0.51	& 0.76	 & 0.04  & -1.35	& 1.4  	 & 0.03  \\
Period     & 5      & 8     & 7     & 6      & 5     & 1      & 7      & 5     \\
  \hline
\end{tabular}
\end{table}
The row ``Period'' gives the time in months between the SNSD
peaks $a_{min}$ and $a_{max}$. As shown in Table~\ref{table1}, 
the time interval between $a_{min}$ and $a_{max}$ is very small, varying from 1 to 8 months.
For some cycles, preceding a weaker cycle, the SNSD peak $a_{max}$ remains negative (cycles 8, 13, and 22).
The largest values of $a_{max}$ exceeding 1 are typical for cycles 1, 2, 7, 17, and 18, which are
followed by stronger cycles. The maximal value of $a_{max}=1.99$ is reached around the peak of cycle 18,
preceding the stronger cycle 19, which is the strongest in the sunspot number record. 
As can be seen in Figure~\ref{fig5}, the SNSD can reach positive values greater than 1 further 
in the declining phase (not around the cycle peak) in cycles (e.g. 11 and 22) that are followed by weaker cycles.
However, this SNSD increase is preceded by prolonged intervals (15 months) of negative SNSD dynamics. 
This gives us reason to assume that there is a direct relation between the SNSD increase (which follows after the SNSD decrease)
and the degree of the SNSD decrease (preceding this SNSD increase). 

\section{Correspondence between the SNSD and sunspot numbers}
In this section we demonstrate that the SNSD dynamics clearly reflect the variations of sunspot numbers,
and even very small SNSD variations are informative indicators of sunspot number dynamics.

As shown in Figure~\ref{fig5}, for some cycles (i.e. 9, 10, 23, and 24) the SNSD peaks $a_{min}$  
are observed 1-3 months earlier than the actual peak of the smoothed sunspot number curve. 
This shift of SNSD peak $a_{min}$ compared to the cycle peak may be a result of the smoothing process itself. 
This process averages several variations of the monthly mean sunspot numbers on time scales smaller than 21 months that are characterized by different peaks. The reduction of the level of smoothing obtained by using $\beta=2$, which corresponds to an effective smoothing interval of only 5 months, keeps these 
variations and makes the smoothed curve closer to the original time series,
but it does not allow us to estimate the trend in the dynamics of the SNSD due to 
their high variability. 

\begin{figure}  
\plotone{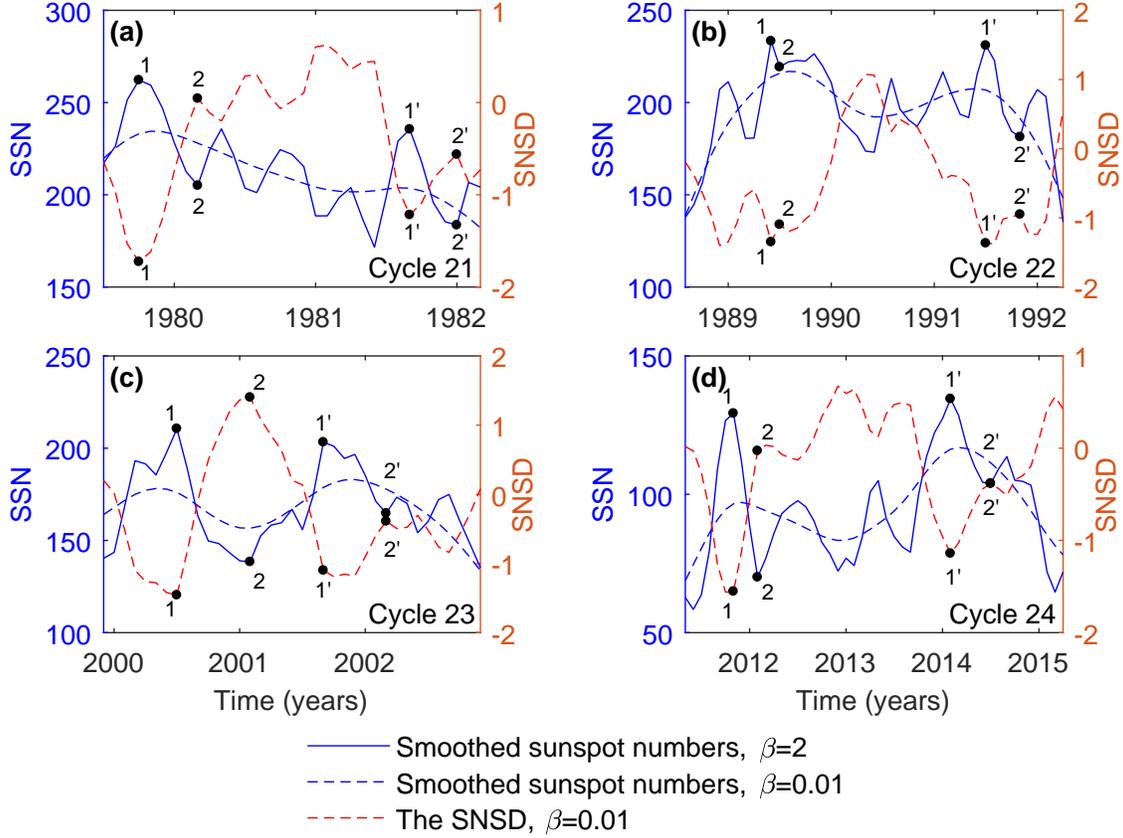}
\caption{Smoothed sunspot numbers for $\beta=2$ (solid blue line, left y axis),
$\beta=0.01$ (dashed blue line, left $y$ axis) and the SNSD for $\beta=0.01$ (dashed red line, right $y$ axis) 
near the maximum of sunspot cycles 21 (a), 22 (b), 23 (c), and 24 (d).}
\label{fig6}
\end{figure}

Figure~\ref{fig6} shows the smoothed sunspot numbers 
for $\beta=2$ (solid blue line, left $y$ axis), $\beta=0.01$ (dashed blue line, left $y$ axis), and 
the SNSD for $\beta=0.01$ (dashed red line, right $y$ axis) near the maximum of sunspot cycles 21 (a), 22 (b),
23 (c), and 24 (d). 
As shown in Figure~\ref{fig6}, the negative SNSD peaks $a_{min}$ for $\beta=0.01$ 
marked by points $1'$ and 1 (dashed red line) correspond to cycle peaks of the smoothed
sunspot curve for $\beta=2$ marked by the point 1 and $1'$ (solid blue line).
The SNSD peaks $a_{max}$ for $\beta=0.01$ marked by 2 and $2'$ (dashed red line)
correspond to the local minima of the smoothed sunspot curve for $\beta=2$ marked 
by the points 2 and $2'$ (blue solid line). 
As can be seen, e.g., from Figure~\ref{fig6}a, the fall of SNSD to a negative peak $a_{min}$  
(points 1 and $1'$) precedes the decrease of sunspot numbers from point 1 ($1'$) to 2 ($2'$).
The growth of SNSD from point 1 ($1'$) to point 2 ($2'$) precedes the increase of sunspot 
numbers from point 2 ($2'$) to the following local maximum.
Moreover, all the local minima of the SNSD curve for $\beta=0.01$ 
(dashed red line) correspond to local maxima of the smoothed sunspot curve for $\beta=2$ 
(blue solid line). These variations, including very small ones, occur in anti-phase
with the variations of sunspot numbers. The SNSD variations (right y axis) in the range 
from -2 to 2 precede sunspot number variations that are significantly larger (more than one order).
Changes in the SNSD trend reflect future changes in the sunspot activity trend.

\section{The relevant indicator for prediction}\label{S-Ind}
We now proceed to make these observations more quantitative, and show how they may be used 
to predict whether the next cycle will be stronger or weaker compared to the current one.
We define the indicator $r_{k}$ that reflects the \textit{relative} intensity 
of the growth process of the SNSD to its peak $a_{max}$ right after reaching the negative peak $a_{min}$
around the cycle maximum:

\begin{equation} \label{Eq_r}
   r_{k}=\frac{a_{max}^{k}}{|a_{min}^{k}|}.
\end{equation}
Here, $k$ is the solar cycle number. 
The parameter $r_k$ basically means we must normalize for each cycle the SNSD peak $a_{max}$
by the corresponding $a_{min}$ value. (Note that $a_{min}$ is negative in all cases).

\begin{figure}  
\plotone{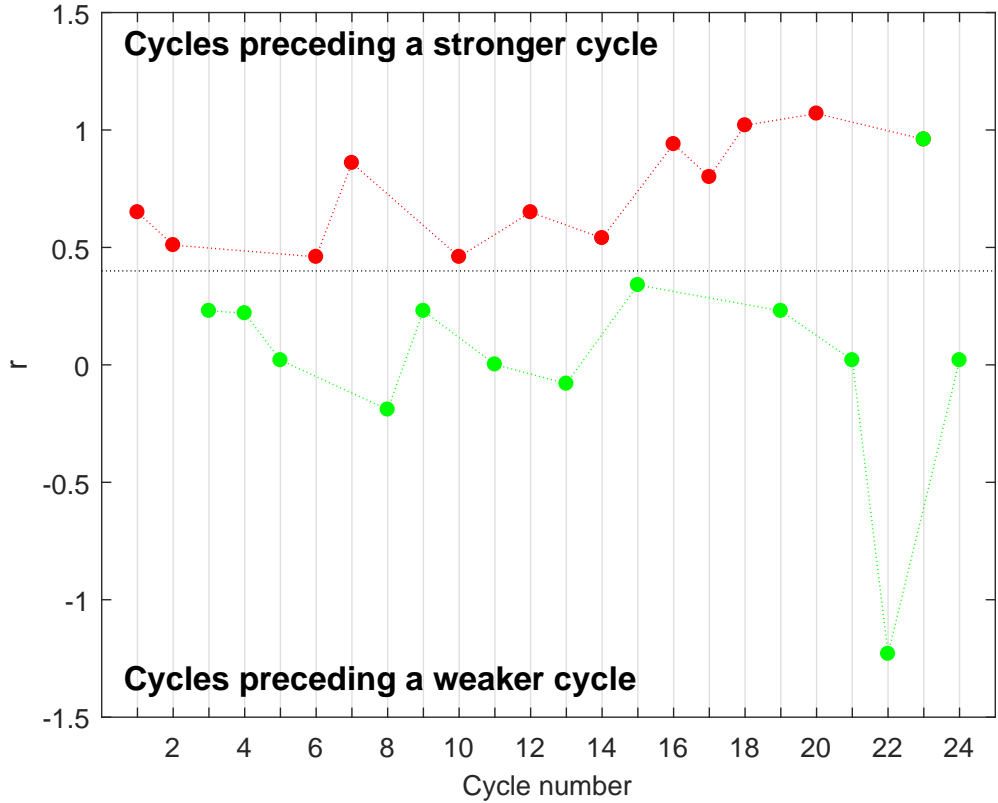}
\caption{The relevant indicator $r_{k}$ for the cycles preceding the stronger cycles (red dots)
and that for the cycles preceding the weaker cycles (green dots).
The horizontal line at $r=0.4$ indicates the empirical threshold 
in the parameter $r_{k}$ used to distinguish between stronger/weaker following cycles.}
\label{fig7}
\end{figure}

Table~\ref{table2} gives the values of $r_{k}$ for cycles 1-24. 
The letters ``S'' or ``W'' in the row ``Strength'' indicate whether the next cycle will be stronger or weaker compared to the current one. 
\begin{table}
\centering
\caption{Relevant indicator $r_{k}$ for cycles 1-24. The row ``Cycle'' gives the current cycle number. 
The row ``Strength'' indicates whether the next cycle will be stronger (S) or weaker (W) compared to the current one.}
\label{table2}
\begin{tabular}{lcccccccc}     
  \hline                         
  \hline                  
Cycle     & 1     & 2     & 3     & 4     & 5     & 6     & 7     & 8     \\
Strength  & S     & S     & W     & W     & W     & S     & S     & W     \\
r         & 0.65  & 0.51  & 0.23  & 0.22  & 0.02  & 0.46  & 0.86  & -0.19 \\
  \hline
Cycle     & 9     & 10    & 11    & 12    & 13    & 14    & 15    & 16    \\
Strength  & W     & S     & W     & S     & W     & S     & W     & S     \\
r         & 0.23  & 0.46  & 0.002 & 0.65  & -0.08 & 0.54  & 0.34  & 0.94  \\
	\hline
Cycle     & 17    & 18    & 19    & 20    & 21    & 22    & 23    & 24    \\
Strength  & S     & S     & W     & S     & W     & W     & W     &       \\
r         & 0.80  & 1.02  & 0.23  & 1.07  & 0.02  & -1.23 & 0.96  & 0.02  \\
  \hline
\end{tabular}
\end{table}
Figure~\ref{fig7} shows the values of $r_{k}$ for the cycles preceding the stronger cycles (red dots) and the values for the cycles preceding the weaker cycles (green dots). 
As is seen in Figure~\ref{fig7}, all the red dots are located well above the level $L=0.4$ depicted 
by the dashed black line, and almost all the green dots are located well below this level. 
The minimal value of $r_{k}$ for the cycles preceding stronger cycles 
is equal to 0.46 (cycles 6 and 10), and the maximal value of $r_{k}$ is equal to 1.07 (cycle 20). 
Except for the cycle 23 ($r_{23}=0.96$, see discussion), the relevant indicator $r_{k}$ for all the cycles 
preceding weaker cycles varies from -1.23 (cycle 22) to 0.34 (cycle 15). 

Thus, we can use the comparison of $r_{k}$  with the empirically determined threshold $L = 0.4$ as 
a criterion to predict whether the amplitude of the next cycle will be larger or smaller compared to the current one.
This approach allows us to provide successful predictions for 22 from 23 tested cycles and proves that the indicator
$r_{k}$ gives a prediction with a probability of 0.96 (22/23) whether the next cycle will be weaker or stronger.
As shown in Table~\ref{table1}, the time interval between the SNSD peaks $a_{min}$ and $a_{max}$ varies
between 1 and 8 months. The prediction can be done about one year after the time of the SNSD peak $a_{max}$, 
as its estimation depends on monthly mean sunspot numbers over the ensuing 10 months. 
The relevant indicator $r_{24}=0.02$ for cycle 24 is significantly smaller than the threshold 
$L=0.4$, therefore, we predict that the next cycle, cycle 25, will be weaker compared to the current cycle, cycle 24.

\section{Optimized moving-average smoothing}
As shown above, the optimized smoothing method introduced in this paper provides a significantly better description 
of the data in terms of both fidelity and smoothness compared to the traditional 13-month running mean (see Figure~\ref{fig3}). However, a disadvantage of the method is that in principal, with the addition of every new data point (i.e. every month), \textit{all} the previous smoothed values need to be recomputed. To circumvent that drawback, in this Section we develop a method for how this optimized smoothing can be adapted to include only a finite interval of data points for the smoothing, thus providing a weighted moving-average on the basis of the proposed optimized smoothing technique.

As discussed in Section 4, the main contribution to the optimized smoothed sunspot number with $\beta=0.01$ comes from 
21 values of the original data series, i.e. by the months $-10$ to $+10$. Thus, we could derive a weighted moving-average considering such a 21 month interval, which would well describe the trend in the data. However, for our present purposes we have to take into account that the SNSD are very sensitive to noise in data, and we thus propose using a smoothing interval of 45 months with changing weights for months -$22$ to $+22$.

The weighted average at every point $i$ with optimized weights is obtained according to the optimality criterion 
given by Equation~(\ref{Eq_J}). Let $R = \left|R_{i-22},R_{i-21},\cdots,R_{i},...,R_{i+21},R_{i+22}\right|^{T}$ 
denote the $45$-dimensional vector of monthly mean sunspot numbers, and 
$\left|\bar{R}_{i-22},\bar{R}_{i-21},\cdots,\bar{R}_{i},\cdots,\bar{R}_{i+21},\bar{R}_{i+22}\right|^{T}$  represent the 
$45$-dimensional vector of monthly smoothed sunspot numbers.
The central point $\bar{R}_{i}$ of the smoothed vector, determined as the weighted sum 
of the 45 components of vector $R$, can be considered as the optimized 45-month moving-average.
The optimized smoothing weights correspond to elements from the middle row of matrix $A^{-1}$ ($45\times 45$-dimensional in this case) multiplied by $\beta$.  Let $C=\left|c_{23,1},c_{23,2},\cdots,c_{23,45}\right|$ denote the middle (23th) row 
of matrix $A^{-1}$ with the elements $c_{23,j}$. Then, at every step $i$ the optimized moving-average
with an interval of 45 months is determined by
\begin{equation} \label{Eq_Sol13}
  \bar{R}_{i}=\beta CR
\end{equation}
The obtained value of the optimized moving-average $\bar{R}_{i}$ belongs to the center of the chosen interval
and gives an estimate of the smoothed sunspot numbers with a delay of 22 months. 
The optimized weights in the 45-month average obtained from vector $\beta C$ depend only 
on the smoothing coefficient $\beta$ and remain constant at every step $i$.

\begin{figure}  
\plotone{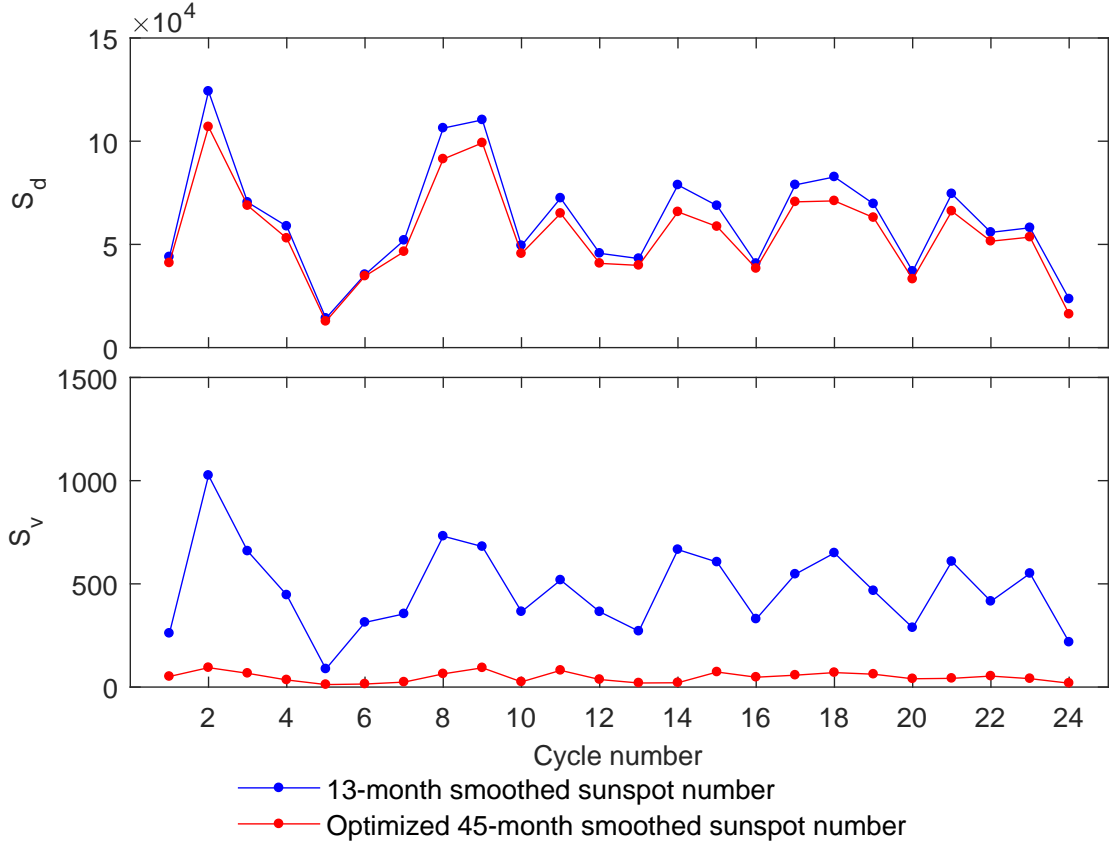}
\caption{Deviation indicator $S_{d}$ (a) and the variability indicator $S_{v}$ (b) for all sunspot cycles 1--24
derived for the 13-month smoothed sunspot numbers (blue), and for the optimized 45-month smoothed sunspot numbers (red).}
\label{fig8}
\end{figure}

Figure~\ref{fig8} shows the deviation and variability indicators $S_d$ (a) and $S_v$ (b) determined
by Equations~(\ref{Eq_Sd}) and (\ref{Eq_Sv}) for the optimized 45-month smoothed sunspot number (red).
The blue lines indicate the values of $S_d$ and $S_v$ for the traditional 13-month running mean. 
The deviations of the obtained indicators $S_{d}$ and $S_{v}$ from those estimated on the basis of
considering all elements of the optimized smoothing method are within 1\%. As can be seen from a comparison of Figure~\ref{fig8} 
and Figure~\ref{fig3}, the deviation and variability indicators $S_d$ and $S_v$ for the optimized 45-month moving-average (red) 
exhibit the same advantages as the full optimized smoothing method, compared to the 13-month smoothed sunspot numbers (blue).

\begin{figure}  
\plotone{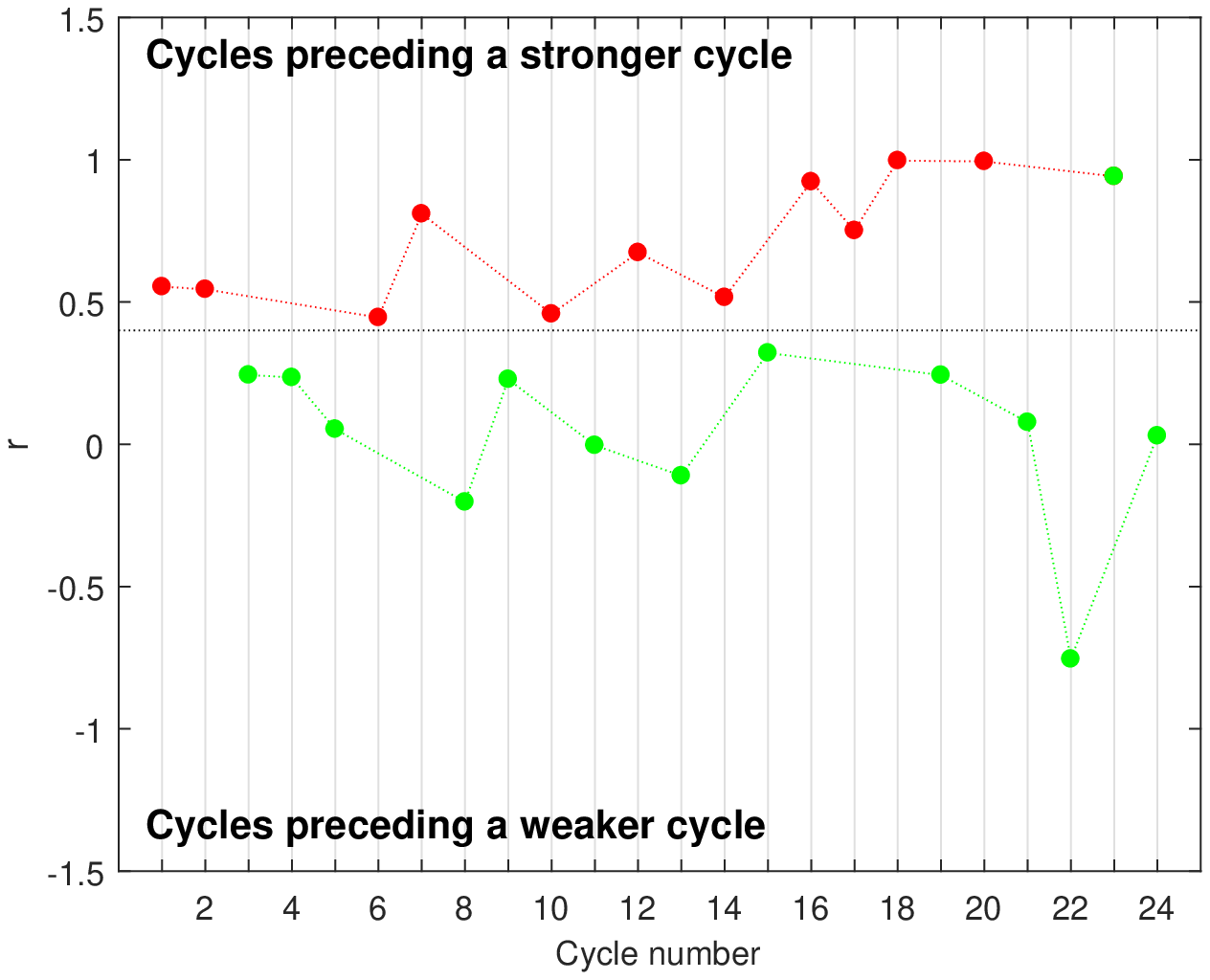}
\caption{Relevant indicator $r_{k}$ constructed on the basis of optimized 45-month smoothed sunspot numbers.
The red dots show $r_{k}$ for the cycles preceding the stronger cycles.
The green dots give $r_{k}$  for the cycles preceding the weaker cycles.
The horizontal line at $r=0.4$ indicates the empirical threshold in the parameter 
$r_{k}$ used to distinguish between stronger/weaker following cycles.}
\label{fig9}
\end{figure}

We repeated the same analysis as in Section 4 but using the optimized 45-month smoothed sunspot data.
Figure~\ref{fig9} shows the relevant indicator $r_{k}$ constructed from the dynamics of SNSD 
estimated on the basis of the optimized 45-month smoothed sunspot numbers.
The red dots show the values of $r_{k}$ for the cycles preceding the stronger cycles,
and the green dots give $r_{k}$ for the cycles preceding the weaker cycles.
The implementation of the optimized 45-month moving-average only slightly changes the peaks $a_{min}$ and $a_{max}$
that are registered at the same time as the peaks in Figure~\ref{fig5}. 
The minimal value of $r_{k}$ for the cycles preceding stronger cycles decreased from 0.46 to 0.45 (cycle 6) when using
the optimized 45-month moving-average. The maximal value of $r_{k}$ for all the cycles preceding weaker cycles decreased
from 0.34 to 0.32. As can be seen from a comparison of Figure~\ref{fig9} and Figure~\ref{fig7}, the relevant indicator $r$, 
used to distinguish between stronger/weaker following cycles, is also slightly changed. But the general trend in Figure~\ref{fig9} 
and the main results (predictive power of $r$) do not change, and the features of the quantified SNSD are kept using 
the optimized 45-month moving-averaged smoothing.

\section{Discussion}
We have demonstrated that the relative growth of SNSD from $a_{min}$ to $a_{max}$ in the early stage 
of the declining phase of the current cycle has predictive potential for the following cycle strength. 
The empirically determined threshold of $L=0.4$ allows a clear distinction between a stronger or weaker 
following cycle for all cycles 1--24 except for cycle 23. In the following we discuss the peculiarities 
of the cycle strength predictions based on the properties of the SNSD dynamics of the current cycle 
characterized by Gnevyshev peaks.

\begin{figure}  
\plotone{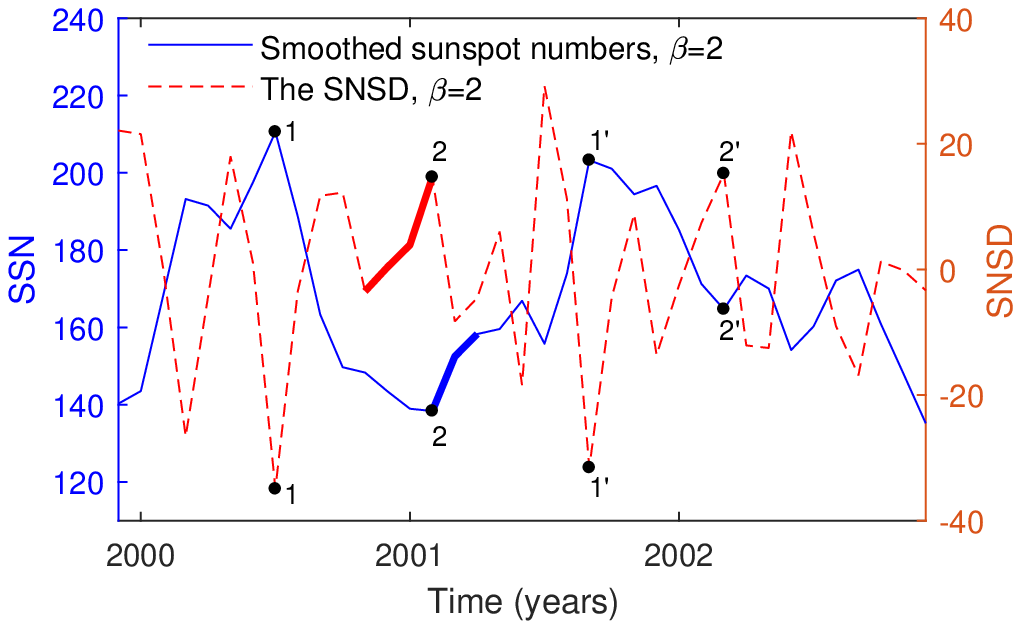}
\caption{The smoothed sunspot numbers for $\beta=2$ (solid blue line, left y axis) and the SNSD for
$\beta=2$ (dashed red line, right y axis) near the maxima of sunspot cycle 23.
The growth of sunspot numbers is preceded by the growth of SNSD.}
\label{fig10}
\end{figure}

Figure~\ref{fig10} shows the smoothed sunspot numbers for $\beta=2$ (solid blue line) near the maxima of sunspot cycle 23. The interval between cycle peaks marked by points  $1'$  and 1 (solid blue line) is 14 months. This is significantly shorter than the intervals for cycles 22 (25 months) and 24 (27 months), which are characterized by a clear Gnevyshev gap.
To determine when the cycle enters to the phase of the second peak, we estimate the SSND for
$\beta=2$ (dashed red line) characterized by highly variable dynamics due to a low level of smoothing; this 
but it allows us to react faster to changes in sunspot number dynamics. Points 1, $1'$, 2, and $2'$ have the
same significance and are located at the same place as in Figure~\ref{fig6}c. As shown in Figure~\ref{fig10}, 
the SNSD oscillations and short-term sunspot number variations occur in anti-phase and the SNSD oscillations
precede those of sunspot numbers. After reaching a local minimum between peaks (point 2, solid
blue line), the sunspot numbers exhibit a tendency for growth (bold solid blue line). This growth
is preceded by the growth of the SNSD (bold solid red line) that is registered 4 months after the first
cycle peak, indicating a very fast entrance of cycle 23 to the phase of the second cycle peak.
For smoothing constant $\beta=0.01$ the smoothing interval is increased (21 months) and the SNSD
growth, preceding the growth of sunspot numbers to the second cycle peak, superimposes on the
SNSD variations around the first cycle peak. Therefore, the SNSD peak $a_{max}$ marked by the point 2
(dashed red line, Figure~\ref{fig6}c) is overestimated. Thus, to exclude the influence of an early entrance of a
cycle on the phase of the second cycle peak, we could potentially employ the analysis of the SNSD
growth from $a_{min}$ to $a_{max}$ around the second peak of cycles characterized by a clear Gnevyshev gap.
The predictive indicators constructed around the second peaks of cycles 22, 23, and 24 
are determined to be $r_{22}=\frac{-0.95}{|-1.37|}=-0.69$, $r_{23}=\frac{-0.39}{|-1.19|}=-0.33$,
and $r_{24}=\frac{-0.38}{|-1.14|}=-0.33$. Comparison of these indicators with the threshold $L = 0.4$ 
indicates that the each following cycle will be weaker.

\section{Conclusions} 
\label{S-conc}
Our study shows a steady relation between the second derivative of the smoothed monthly mean sunspot numbers
around the peak of the current cycle and the strength of the ensuing cycle.
The second derivatives of sunspot numbers were estimated as second differences 
of smoothed sunspot numbers (SNSD); this is a quantity that is closely related to the curvature 
of the sunspot series. The noisy record in sunspot numbers was 
processed using an optimized smoothing technique for the monthly mean sunspot numbers, in order to provide
a level of smoothing that is sufficient to reliably derive the SNSDs and to study their dynamics.
The proposed optimized smoothing technique is based on finding a balance between the fidelity 
to the data and the smoothness of the approximating curve. It is demonstrated to provide 
a better approximation of the monthly mean sunspot numbers than the traditional 13-month running mean.

Based on the optimized smoothing method introduced in this paper, we have also created 
a weighted moving-average method. Here, we have applied it over a smoothing interval of $\pm22$ months, 
in order to provide a reliable estimate of the SNSD, and we have shown that our results obtained remain 
stable with this method. However, we note that for more general descriptions of the solar cycle, 
this method can be also applied in the same way to a range of $\pm6$ months. 
In this way it corresponds to the traditional 13-month running mean but provides 
a better description of the sunspot series in terms of data fidelity and smoothness.

We demonstrated that after reaching a negative peak $a_{min}$ around a cycle peak,
the ensuing growing trend of the SNSD exhibits stable differences for cycles followed by stronger cycles and for those followed by weaker cycles. 
A relevant indicator was constructed from the dynamics of the SNSD to determine whether the strength of the next cycle 
will be stronger or weaker compared to the current one.
The proposed approach has been successfully demonstrated for 22 out of 23 cycles studied. 
We may speculate that the dynamics of the SNSD variations reflect the nature of interaction 
of the poloidal and toroidal components of the large-scale solar magnetic field, when the evolution 
of the toroidal fields due to the alpha-effect gradually leads to an increase of the poloidal field component, 
but with opposite polarity. This finding has a physical basis in dynamo models \citep{Nandy2002} that involve 
the poloidal magnetic field from the current cycle as a precursor of the next cycle \citep{{Schatten2005},{Svalgaard2005}}. 
In a recent study, \citet{CameronSchlussler2015} have demonstrated that the magnetic flux observed in sunspot groups is effectively 
driven by the polar fields. They showed that the polar fields of the preceding cycle are the dominant source 
of the net toroidal flux, from which the sunspot groups of the actual cycle originate. These findings provide 
a firm physical basis and understanding for why precursor methods do actually work. 

The unique unsuccessful prediction based on the relevant indicator constructed in the case of cycle 23 might 
be related to the superposition of variations around the first and second cycle peaks.
The construction of a relevant indicator after the second peak of cycles that reveal a clear Gnevyshev gap
could be potentially employed to exclude the influence of the early entrance of a cycle on the phase of the second cycle peak.

Thus, we have shown that the relevant indicator constructed for the current cycle gives a prediction with a probability of 0.96 
for whether the next cycle will be weaker or stronger. The SNSD characteristics required to determine the relevant predictive indicator
($a_{min}$,$a_{max}$) are located within an interval of 1--8 months. The prediction additionally requires about 2 years, 
as the estimation of the relevant indicator depends on monthly mean sunspot numbers over the ensuing 22 months. 
This means that within 24--31 months after reaching the maximum of a cycle, a prediction for the following cycle can be made.
On the basis of this indicator we predict that the next sunspot cycle, cycle 25, will be weaker than the current one.

\begin{acknowledgements}
The authors acknowledge the team of WDC-SILSO at the Royal Observatory of Belgium (ROB) for the sunspot number data sets.	
We thank the referee for very valuable and insightful comments on this study.
\end{acknowledgements}

%\bibliography{My_References}

\begin{thebibliography}{}
\expandafter\ifx\csname natexlab\endcsname\relax\def\natexlab#1{#1}\fi
\providecommand{\url}[1]{\href{#1}{#1}}

\bibitem[{{Bohlmann}(1899)}]{Bohlmann1899}
{Bohlmann}, G. 1899, Math. Phys. Klasse, 3, 260

\bibitem[{{Braj{\v s}a} {et~al.}(2009){Braj{\v s}a}, {W{\"o}hl}, {Hanslmeier},
  {Verbanac}, {Ru{\v z}djak}, {Cliver}, {Svalgaard}, \& {Roth}}]{Braja2009}
{Braj{\v s}a}, R., {W{\"o}hl}, H., {Hanslmeier}, A., {et~al.} 2009, Cent. Eur.
  Astrophys. Bull., 33, 95

\bibitem[{{Brown}(1963)}]{Brown1963}
{Brown}, R.~G. 1963, {Smoothing Forecasting and Prediction in Discrete Time Series} (Englewood Cliffs, NJ: Prentice-Hall), 468

\bibitem[{{Cameron} \& {Sch{\"u}ssler}(2007)}]{CameronSchlussler2007}
{Cameron}, R., \& {Sch{\"u}ssler}, M. 2007, Astrophys. J., 659, 801

\bibitem[{{Cameron} \& {Sch{\"u}ssler}(2008)}]{CameronSchlussler2008}
---. 2008, Astrophys. J., 685, 1291

\bibitem[{{Cameron} \& {Sch{\"u}ssler}(2015)}]{CameronSchlussler2015}
---. 2015, Science, 347, 1333

\bibitem[{{Clette} {et~al.}(2016){Clette}, {Lef{\`e}vre}, {Cagnotti},
  {Cortesi}, \& {Bulling}}]{Clette2016}
{Clette}, F., {Lef{\`e}vre}, L., {Cagnotti}, M., {Cortesi}, S., \& {Bulling},
  A. 2016, \solphys, arXiv:1507.07803

\bibitem[{{Conway}(1998)}]{Conway1998}
{Conway}, A.~J. 1998, New Astron. Rev., 42, 343

\bibitem[{{Feynman}(1982)}]{Feynman1982}
{Feynman}, J. 1982, \jgr, 87, 6153

\bibitem[{{Gnevyshev}(1967)}]{Gnevyshev1967}
{Gnevyshev}, M.~N. 1967, \solphys, 1, 107

\bibitem[{{Gonzalez} \& {Schatten}(1988)}]{GonzalezSchatten1988}
{Gonzalez}, G., \& {Schatten}, K.~H. 1988, \solphys, 114, 189

\bibitem[{{Hathaway}(2009)}]{Hathaway2009}
{Hathaway}, D.~H. 2009, \ssr, 144, 401

\bibitem[{{Hathaway}(2010)}]{Hathaway2010}
---. 2010, Living Rev. Solar Phys., 7, 1

\bibitem[{{Hathaway} {et~al.}(1999){Hathaway}, {Wilson}, \&
  {Reichmann}}]{Hathaway1999}
{Hathaway}, D.~H., {Wilson}, R.~M., \& {Reichmann}, E.~J. 1999, \jgr, 104,
  22375

\bibitem[{{Hurst}(1970)}]{Hurst1970}
{Hurst}, J.~M. 1970, {The Profit Magic of Stock Transaction Timing} (Englewood Cliffs, NJ: Prentice-Hall), 223

\bibitem[{{Kane}(2008)}]{Kane2008}
{Kane}, R.~P. 2008, \solphys, 248, 203

\bibitem[{{Lantos}(2006)}]{LantosSkewness2006}
{Lantos}, P. 2006, \solphys, 236, 199

\bibitem[{{Mu{\~n}oz-Jaramillo} {et~al.}(2012){Mu{\~n}oz-Jaramillo}, {Sheeley},
  {Zhang}, \& {DeLuca}}]{MunozJaramillo2012}
{Mu{\~n}oz-Jaramillo}, A., {Sheeley}, N.~R., {Zhang}, J., \& {DeLuca}, E.~E.
  2012, \apj, 753, 146

\bibitem[{{Nandy} \& {Choudhuri}(2002)}]{Nandy2002}
{Nandy}, D., \& {Choudhuri}, A.~R. 2002, Science, 296, 1671

\bibitem[{{Ohl} \& {Ohl}(1979)}]{OhlOhl1979}
{Ohl}, A.~I., \& {Ohl}, G.~I. 1979,
Marshall Space Flight Center Solar-Terrest.
Predictions Proc. Vol. 2 (NASA), 258


\bibitem[{{Petrovay}(2010)}]{Petrovay2010}
{Petrovay}, K. 2010, Liv. Rev. Sol. Phys., 7, 6

\bibitem[{{Podladchikova} {et~al.}(2008){Podladchikova}, {Lefebvre}, \& {Van
  der Linden}}]{PodladchikovaLefebvreLinden2008}
{Podladchikova}, T., {Lefebvre}, B., \& {Van der Linden}, R. 2008, J. Atm.
  Sol-Terr. Phys., 70, 277

\bibitem[{{Podladchikova} \& {van der Linden}(2011)}]{Podladchikova2011}
{Podladchikova}, T., \& {van der Linden}, R. 2011, Journal of Space Weather and
  Space Climate, 1, A260000

\bibitem[{{Ramaswamy}(1977)}]{Ramaswamy1977}
{Ramaswamy}, G. 1977, \nat, 265, 713

\bibitem[{{Rice}(1981)}]{Rice81}
{Rice}, J.~R. 1981, {Matrix Computations and Mathematical Software} (New York,  NY: McGraw-Hill), 248

\bibitem[{{Schatten} {et~al.}(1996){Schatten}, {Myers}, \&
  {Sofia}}]{Schatten1996}
{Schatten}, K., {Myers}, D.~J., \& {Sofia}, S. 1996, \grl, 23, 605

\bibitem[{{Schatten}(2005)}]{Schatten2005}
{Schatten}, K.~H. 2005, \grl, 32, L21106

\bibitem[{{Schatten} {et~al.}(1978){Schatten}, {Scherrer}, {Svalgaard}, \&
  {Wilcox}}]{SchattenDynamo1978}
{Schatten}, K.~H., {Scherrer}, P.~H., {Svalgaard}, L., \& {Wilcox}, J.~M. 1978,
  \grl, 5, 411

\bibitem[{{Schatten} \& {Sofia}(1987)}]{SchattenSofia1987}
{Schatten}, K.~H., \& {Sofia}, S. 1987, \grl, 14, 632

\bibitem[{{Svalgaard} {et~al.}(2005){Svalgaard}, {Cliver}, \&
  {Kamide}}]{Svalgaard2005}
{Svalgaard}, L., {Cliver}, E.~W., \& {Kamide}, Y. 2005, \grl, 32, 1104

\bibitem[{{Thompson}(1993)}]{Thompson1993}
{Thompson}, R.~J. 1993, \solphys, 148, 383

\bibitem[{Wang \& Sheeley(2009)}]{WangSheeley2009}
Wang, Y.-M., \& Sheeley, N.~R. 2009, The Astrophysical Journal Letters, 694,
  L11

\bibitem[{Weinert(2007)}]{Weinert2007}
Weinert, H.~L. 2007, Comput. Stat. Data Anal., 52, 959

\bibitem[{{Whittaker}(1923)}]{Whittaker1923}
{Whittaker}, E.~T. 1923, in Proc. Edinburgh Math. Soc 41 (Cambridge: Cambridge Univ. Press), 63

\bibitem[{{Wilson} {et~al.}(1998){Wilson}, {Hathaway}, \&
  {Reichmann}}]{WilsonHathawayReichmann1998}
{Wilson}, R.~M., {Hathaway}, D.~H., \& {Reichmann}, E.~J. 1998, \jgr, 103, 6595

\end{thebibliography}

\end{document}